\documentclass [12pt]{article}
\pdfoutput=1

\usepackage{amsmath}
\usepackage{amsfonts}
\usepackage{amscd}
\usepackage{amsthm}
\usepackage{setspace}

\usepackage{graphicx}
\usepackage{authblk}
\usepackage{caption}
\usepackage{ytableau}
\usepackage{mathtools}

\def\Q{\mathbb{Q}}

\begin{document}

\begin{titlepage}

\begin{flushright}
KEK-TH-2249
\end{flushright}

\vskip 3.0cm

\begin{center}

{\large JT gravity and the asymptotic Weil--Petersson volume}

\vskip 1.2cm

Yusuke Kimura$^1$ 
\vskip 0.6cm
{\it $^1$KEK Theory Center, Institute of Particle and Nuclear Studies, KEK, \\ 1-1 Oho, Tsukuba, Ibaraki 305-0801, Japan}
\vskip 0.4cm
E-mail: kimurayu@post.kek.jp

\vskip 2cm
\abstract{A path integral in Jackiw--Teitelboim (JT) gravity is given by integrating over the volume of the moduli of Riemann surfaces with boundaries, known as the ``Weil--Petersson volume,'' together with integrals over wiggles along the boundaries. The exact computation of the Weil--Petersson volume $V_{g,n}(b_1, \ldots, b_n)$ is difficult when the genus $g$ becomes large. We utilize two partial differential equations known to hold on the Weil--Petersson volumes to estimate asymptotic behaviors of the volume with two boundaries $V_{g,2}(b_1, b_2)$ and the volume with three boundaries $V_{g,3}(b_1, b_2, b_3)$ when the genus $g$ is large. Furthermore, we present a conjecture on the asymptotic expression for general $V_{g,n}(b_1, \ldots, b_n)$ with $n$ boundaries when $g$ is large.}  

\end{center}
\end{titlepage}

\tableofcontents
\section{Introduction}
\par Jackiw--Teitelboim (JT) gravity \cite{Teitelboim1983, Jackiw1984} is a two-dimensional (2d) quantum gravitational theory. In the case of bosonic JT gravity, the path integral is essentially determined \cite{SSS2019} from the volume of the moduli of hyperbolic Riemann surfaces, otherwise known as the ``Weil--Petersson volume,'' and the path integrals over ``wiggles'' along boundaries of Riemann surfaces \cite{Jensen2016, MSY2016, EMV2016} when these surfaces have a boundary. Schwarzian theory \cite{Kitaevtalk, MS2016, Kitaev2017} controls the wiggles. Therefore, Sachdev--Ye--Kitaev (SYK) models \cite{Sachdev1992, Kitaevtalk, Kitaev2017} are related to JT gravity, as the SYK models are described by the one-dimensional Schwarzian theory at low energies. Recent progress of JT gravity can be found, e.g., in \cite{Blommaert2019, SSS2019, Maldacena2019, Mertens2019, Iliesiu2019, Cotler2019, Moitra2019, StanfordWitten2019, Gross2019, Okuyama2019, Penington2019, Almheiri2019, Marolf2020, OkuyamaSakai2020multi, Witten2020}.
\par The genus $g$ partition function with $n$ boundaries, $Z_{g,n}(\beta_1, \ldots, \beta_n)$, is given by an integral over a function involving the Weil--Petersson volume of the moduli of Riemann surfaces of genus $g$ with $n$ boundaries \cite{SSS2019}.

\vspace{5mm}

\par A Hermitian-matrix integral computes the JT gravity path integral \cite{SSS2019}. This is owing to the fact that Mirzakhani's recursion relation \cite{Mirzakhani2007}, which holds for the Weil--Petersson volumes of the moduli of the Riemann surface on the JT gravity side, corresponds \cite{Eynard2007} to the topological recursion of Eynard and Orantin \cite{Eynard2007tp}, which yields the genus expansion of a Hermitian-matrix integral as discussed in \cite{Eynard2004}.

\vspace{5mm} 

\par In JT gravity, connected correlators are given as the sums over genus $g$ JT path integrals \cite{SSS2019}:
\begin{equation}
<Z(\beta_1)\ldots Z(\beta_n)>_{c} \simeq \sum^{\infty}_{g=0} \frac{Z_{g}(\beta_1, \ldots, \beta_n)}{(e^{S_0})^{2g-2+n}}. 
\end{equation}
The genus $g$ partition function with $n$ boundaries, $Z_{g}(\beta_1, \ldots, \beta_n)$ is an integral over a function involving the Weil--Petersson volume $V_{g,n}$ \cite{SSS2019}
\begin{equation}
Z_{g}(\beta_1, \ldots, \beta_n) = \alpha^n \prod_{i=1}^n \int_0^{\infty} b_i db_i V_{g,n}(b_1, \ldots, b_n) \prod^n_{j=1} Z^{\rm trumpet}_{\rm Sch} (\beta_j, b_j).
\end{equation}
Therefore, connected correlators and path integrals in JT gravity can be computed when the Weil--Petersson volume $V_{g,n}(b_1, \ldots, b_n)$ is known (together with Schwarzian theory along the boundaries). For this reason, the evaluation of the Weil--Petersson volume $V_{g,n}(b_1, \ldots, b_n)$ has a physical importance. 
\par In this work, we aim to estimate the asymptotic behavior of the Weil--Petersson volume $V_{g,2}(b_1, b_2)$ with genus $g$ and two boundaries of geodesic lengths $b_1$ and $b_2$ when $g$ is large. We also estimate the asymptotic behavior of the Weil--Petersson volume with three boundaries, $V_{g,3}(b_1, b_2, b_3)$, when $g$ is large. Furthermore, we extend these results to propose a conjecture concerning an asymptotic expression for $V_{g,n}(b_1, \ldots, b_n)$. This study focuses on bosonic JT gravity.
\par In mathematics, Mirzakhani's recursion relation \cite{Mirzakhani2007} is known for the Weil--Petersson volume $V_{g,n}(b_1, \ldots, b_n)$, which provides a method for computing the Weil--Petersson volume recursively. Although computational results are known up to some finite genus $g$ and up to some finite number of boundaries $n$, the computation becomes difficult with increasing $g$. In this note, we study the asymptotic behavior of the Weil--Petersson volume $V_{g,n}(b_1, \ldots, b_n)$ when the genus $g$ of Riemann surfaces becomes large \footnote{There are conjectures by Zograf \cite{Zograf2008} for $V_{g,n}$.}. 
\par Mirzakhani's recursion relation expresses the derivative of $b_1\, V_{g,n}$ in terms of $V_{g', n'}$, where $2g'+n'$ is less than $2g+n$. However, this expression necessarily involves $V_{g-1,n+1}$. For this reason, when $g$ is large, it is considerably difficult to deduce the asymptotic expression for $V_{g, n+1}$ from that of $V_{g,n}$ recursively. To resolve this difficulty, we utilize partial differential equations \cite{DoNorbury, Do2008} that hold for the Weil--Petersson volume $V_{g,n}$. This approach yields $V_{g,n+1}$ from $V_{g,n}$ up to the leading order in $g$, when the genus $g$ is sufficiently large. This approach might be useful in evaluating the JT path integral and connected correlators in bosonic JT gravity. 

\vspace{5mm}

\par This report is structured as follows. Section \ref{sec2} summarizes our strategy to estimate the asymptotic behavior of the Weil--Petersson volume $V_{g,n}$ ($n\ge 2$) when the genus $g$ is large. In Section \ref{sec3.1}, we provide an explicit calculation for $V_{g,2}(b_1, b_2)$ when the genus $g$ is large. A leading term in $g$ is obtained, and a subleading correction $\sim 1/g$ is also discussed. In Section \ref{sec3.2}, we estimate the asymptotic expression for $V_{g,3}(b_1, b_2, b_3)$, namely the Weil--Petersson volume with genus $g$ and three geodesic boundaries of lengths $b_1, b_2, b_3$, when $g$ is large. We also provide a conjecture on the asymptotic expression for $V_{g,n}(b_1, \ldots, b_n)$ when $g$ is large. In Section \ref{sec4} we comment on the asymptotic Weil--Petersson volume when the geodesic lengths $b_1, \ldots, b_n$ are large, and we mention a related question. When $n=1$, the Weil--Petersson volume in this region was predicted in \cite{SSS2019}. Section \ref{sec5} closes with concluding remarks and an outlook on some open problems. 

\section{Summary of our strategy to estimate \\ asymptotic $V_{g,n}(b_1, \ldots, b_n)$}
\label{sec2}
As noted in the Introduction, we analyze the asymptotic behavior of the Weil--Petersson volume $V_{g,n}(b_1, \ldots, b_n)$ when the genus $g$ is large. Here, $V_{g,n}(b_1, \ldots, b_n)$ denotes the Weil--Petersson volume of the moduli of genus $g$ Riemann surfaces with $n$ boundaries of geodesic lengths $b_1, \ldots, b_n$. The evaluation of the Weil--Petersson volume enables the computation of the JT path integral and of the connected correlators \cite{SSS2019}. 

\par In performing this analysis, if the asymptotic expression for the volume $V_{g,n}$ is known for a large enough $g$, can the asymptotic expression be deduced for the volume $V_{g, n+1}$? 
\par In principle, Mirzakhani's recursion formula \cite{Mirzakhani2007} provides a method for computing $V_{g,n}$ for every $g$ and $n$, starting from $V_{0,3}$ and $V_{1,1}$, recursively. However, computing $V_{g,n}$ precisely and directly from the recursion formula is considerably difficult when $g$ is large. When the asymptotic expression for $V_{g,n}$ is concerned, there is the additional difficulty of estimating $V_{g,n+1}$ from $V_{g,n}$ from the recursion formula. Mirzakhani's recursion formula expresses $\partial_{b_1} b_1V_{g,n+1}(b_1, \ldots, b_{n+1})$ as the sum of the integrals of $V_{g, n}$ times a function, the products $V_{g_1, n_1} V_{g_2, n_2}$ times a function (where $g_1$ and $g_2$ add up to $g$, and $n_1$ and $n_2$ add to $n+2$), and $V_{g-1, n+2}$ times a function. Therefore, to estimate the asymptotic behavior of $V_{g,n+1}$ from $V_{g,n}$ via the recursion formula, one needs to know the asymptotic behavior of $V_{g-1, n+2}$ where $g$ is large. 
\par To avoid this difficulty, our approach utilizes partial differential equations that hold for $V_{g,n}(b_1, \ldots, b_n)$, as deduced in \cite{DoNorbury, Do2008}. The Weil--Petersson volumes satisfy the following partial differential equations \cite{DoNorbury, Do2008}:
\begin{eqnarray}
\label{DN equations in 2}
\partial_{n+1} V_{g, n+1}(b_1, \ldots, b_n, 2\pi i) = & 2\pi i (2g-2+n)\, V_{g,n} (b_1, \ldots, b_n) \\ \nonumber
\partial^2_{n+1} V_{g, n+1}(b_1, \ldots, b_n, 2\pi i) = & \sum^n_{j=1}b_j \partial_j V_{g,n}(b_1, \ldots, b_n)-(4g-4+n) V_{g,n}(b_1, \ldots, b_n). 
\end{eqnarray}
Here, $\partial_j$ represents the derivative with respect to $b_j$, where $j=1, \ldots, n+1$. 
\par Because we focus on the asymptotic expressions for $V_{g,n}$, we impose the following asymptotic conditions on the genus $g$ and the geodesic lengths of the boundaries, $b_1, \ldots, b_n$:
\begin{equation}
\label{conditions in 2}
g>> 1 \hspace{1 cm} g >> b_i \hspace{2mm} (i=1, \ldots, n).
\end{equation}
Then, we may replace the second differential equation in (\ref{DN equations in 2}) with the following reduced equation under the asymptotic conditions (\ref{conditions in 2}):
\begin{equation}
\label{reduced 2nd equation in 2}
\partial^2_{n+1} V_{g, n+1}(b_1, \ldots, b_n, 2\pi i) = -(4g-4+n) V_{g,n}(b_1, \ldots, b_n).
\end{equation}
\par The partial differential equations (\ref{DN equations in 2}) impose highly nontrivial constraints on the asymptotic expressions for the Weil--Petersson volumes $V_{g,n}$. For example, when $n=0$, the first differential equation in (\ref{DN equations in 2}) becomes \cite{DoNorbury, Do2008}
\begin{equation}
\label{equation for Vg1 in 2}
V'_{g,1}(2\pi i) = 2\pi i(2g-2)\, V_{g,0}.
\end{equation}
Asymptotic expressions for $V_{g,0}$ and $V_{g,1}(b)$ were predicted in \cite{SSS2019} from the matrix-integral analysis using a contour integral for large $g$. One can verify that, when $g>>1$ and $g>>b$, the formulas for $V_{g,0}$ and $V_{g,1}(b)$ in \cite{SSS2019} satisfy the equation (\ref{equation for Vg1 in 2}). This provides a consistency check of the asymptotic expressions for $V_{g,0}$ and $V_{g,1}(b)$ predicted in \cite{SSS2019}. 
\par Here, we estimate $V_{g,2}(b_1, b_2)$ when $g$ is large by applying (\ref{DN equations in 2}), (\ref{reduced 2nd equation in 2}) to the asymptotic expression for $V_{g,1}(b)$ obtained in \cite{SSS2019}. The differential equations in (\ref{DN equations in 2}) and (\ref{reduced 2nd equation in 2}) are highly effective for estimating $V_{g,2}$ when this is asymptotic in $g$. Furthermore, applying (\ref{DN equations in 2}), (\ref{reduced 2nd equation in 2}) to the deduced asymptotic expression for $V_{g,2}(b_1, b_2)$, we also estimate $V_{g,3}(b_1, b_2, b_3)$ when $g$ is large. The iteration of this process leads us to a conjecture on the asymptotic expression for $ V_{g,n}(b_1, \ldots, b_n)$ when $g$ is large. 

\par One can compute large genus contributions to the JT path integral and the connected correlators from the deduced expressions. 

\section{Asymptotic behavior of the Weil--Petersson volume \\ $V_{g,n}(b_1, \ldots, b_n)$ with large genus $g$}
\label{sec3}

\subsection{Asymptotic $V_{g,2}(b_1, b_2)$}
\label{sec3.1}
We estimate the asymptotic expression for $V_{g,2}(b_1, b_2)$ when the genus $g$ is large, using the partial differential equations (\ref{DN equations in 2}) deduced in \cite{DoNorbury, Do2008}. When the Riemann surface has two boundaries, the differential equations (\ref{DN equations in 2}) become
\begin{eqnarray}
\label{DN differential for g2 in 3.1}
\partial_2 V_{g, 2}(b_1, 2\pi i) = & 2\pi i (2g-1)\, V_{g,1} (b_1) \\ \nonumber
\partial^2_2 V_{g, 2}(b_1, 2\pi i) = & b_1 \partial_1 V_{g,1}(b_1)-(4g-3) V_{g,1}(b_1). 
\end{eqnarray}
As stated in Section \ref{sec2}, we impose asymptotic conditions on the genus $g$ and geodesic lengths $b_1, b_2$ as follows:
\begin{equation}
\label{condition g2 in 3.1}
g>> 1 \hspace{1 cm} g >> b_1, b_2.
\end{equation}
The second equation in (\ref{DN differential for g2 in 3.1}) becomes reduced to the following equation under these conditions:
\begin{equation}
\label{reduced 2nd equation in 3.1}
\partial^2_2 V_{g, 2}(b_1, 2\pi i) = -(4g-3) V_{g,1}(b_1).
\end{equation}
\par From a physical argument on 2d topological gravity under the condition $g>>1$ and $g>>b_1$, the asymptotic form of $V_{g,1}(b_1)$ is predicted to become \cite{SSS2019}
\begin{equation}
\label{Vg1 in 3.1}
V_{g,1}(b_1) \sim \frac{4\, (4\pi^2)^{2g-\frac{3}{2}}}{(2\pi)^\frac{3}{2}} \Gamma(2g-\frac{3}{2}) \frac{{\rm sinh}(\frac{b_1}{2})}{b_1}.
\end{equation}
From this $V_{g,1}(b_1)$, we estimate the asymptotic expression for $V_{g,2}(b_1, b_2)$ when the conditions (\ref{condition g2 in 3.1}) are satisfied. It is well known in mathematics that any $V_{g,n}(b_1, \ldots, b_n)$ is symmetric about $b_1, \ldots, b_n$. This implies in particular that $V_{g,2}$ must be symmetric about $b_1$ and $b_2$. Utilizing this symmetry property of the Weil--Petersson volume and the first equation in (\ref{DN differential for g2 in 3.1}), one is naturally led to considering the following asymptotic expression \footnote{Because the Weil--Petersson volume $V_{g,n}(b_1, \ldots, b_n)$ is given as an integral of the exterior power of the Weil--Petersson symplectic form over the moduli space of complete hyperbolic surfaces of the genus $g$ with $n$ boundaries, one naturally expects that the asymptotic expression for the Weil--Petersson volume is given in terms of hyperbolic functions in the large $g$ limit with $g>>b_1, b_2$. Other seemingly straightforward candidate functions (in terms of hyperbolic functions symmetric under exchange of $b_1$ and $b_2$) do not satisfy equations (\ref{DN differential for g2 in 3.1}) to the leading order in $g$.} for $V_{g,2}(b_1, b_2)$:
\begin{equation}
\label{complex Vg2 in 3.1}
V_{g,2}(b_1, b_2) \sim 4\sqrt{\frac{2}{\pi}}(4\pi^2)^{2g-1}\Gamma(2g-\frac{3}{2})(2g-1)\, \frac{{\rm sinh}(\frac{b_1}{2})}{b_1}\frac{{\rm sinh}(\frac{b_2}{2})}{b_2}.
\end{equation}
When $g$ is large ($g >>1$), one can replace $\Gamma(2g-\frac{3}{2})(2g-1)$ with $\Gamma(2g-\frac{1}{2})$ because 
\begin{equation}
\frac{\Gamma(2g-\frac{3}{2})(2g-1)}{\Gamma(2g-\frac{1}{2})}=1+\frac{1}{4g-3} \rightarrow 1
\end{equation}
as $g$ tends toward infinity. Thus, a natural asymptotic expression for $V_{g,2}(b_1, b_2)$ under the conditions (\ref{condition g2 in 3.1}) is
\begin{equation}
\label{Vg2 in 3.1}
V_{g,2}(b_1, b_2) \sim 4\sqrt{\frac{2}{\pi}}(4\pi^2)^{2g-1}\Gamma(2g-\frac{1}{2}) \, \frac{{\rm sinh}(\frac{b_1}{2})}{b_1}\frac{{\rm sinh}(\frac{b_2}{2})}{b_2}.
\end{equation}
The coefficient is chosen to satisfy the first equation in (\ref{DN differential for g2 in 3.1}).

\vspace{5mm}

\par Confirming whether the asymptotic expression (\ref{Vg2 in 3.1}) actually satisfies the reduced differential equation (\ref{reduced 2nd equation in 3.1}) yields a nontrivial check. One can confirm that (\ref{Vg2 in 3.1}) indeed satisfies equation (\ref{reduced 2nd equation in 3.1}) as follows: when our expression (\ref{Vg2 in 3.1}) is substituted into the left-hand side of (\ref{reduced 2nd equation in 3.1}), we obtain
\begin{eqnarray}
\partial^2_2 V_{g,2}(b_1, 2\pi i) & \sim 4\sqrt{\frac{2}{\pi}}(4\pi^2)^{2g-1}\Gamma(2g-\frac{1}{2}) \, \frac{{\rm sinh}(\frac{b_1}{2})}{b_1}\partial^2_2 \frac{{\rm sinh}(\frac{b_2}{2})}{b_2}|_{b_2=2\pi i} \\ \nonumber
& =-4\sqrt{\frac{2}{\pi}}(4\pi^2)^{2g-2}\Gamma(2g-\frac{1}{2}) \, \frac{{\rm sinh}(\frac{b_1}{2})}{b_1}.
\end{eqnarray}
Substituting (\ref{Vg1 in 3.1}) into the right-hand side of (\ref{reduced 2nd equation in 3.1}) yields
\begin{eqnarray}
-(4g-3)\, V_{g,1}(b_1) & = -\frac{4\, (4\pi^2)^{2g-\frac{3}{2}}}{(2\pi)^\frac{3}{2}} \Gamma(2g-\frac{3}{2})(4g-3) \frac{{\rm sinh}(\frac{b_1}{2})}{b_1} \\ \nonumber
& =-4\sqrt{\frac{2}{\pi}}(4\pi^2)^{2g-2} \frac{4g-3}{2} \Gamma(2g-\frac{3}{2})\frac{{\rm sinh}(\frac{b_1}{2})}{b_1} \\ \nonumber
& =-4\sqrt{\frac{2}{\pi}}(4\pi^2)^{2g-2} \Gamma(2g-\frac{1}{2})\frac{{\rm sinh}(\frac{b_1}{2})}{b_1},
\end{eqnarray}
because $2 \, \frac{4\, (4\pi^2)^{2g-\frac{3}{2}}}{(2\pi)^\frac{3}{2}}=4\sqrt{\frac{2}{\pi}}(4\pi^2)^{2g-2}$, and $\frac{4g-3}{2} \Gamma(2g-\frac{3}{2})=\Gamma(2g-\frac{1}{2})$. This confirms that the left- and right-hand sides of (\ref{reduced 2nd equation in 3.1}) are equal when $V_{g,2}(b_1, b_2)$ is expressed as (\ref{Vg2 in 3.1}). We thus confirmed that (\ref{Vg2 in 3.1}) satisfies the reduced equation (\ref{reduced 2nd equation in 3.1}). 
\par The expression for $V_{g,2}(b_1, b_2)$ (\ref{Vg2 in 3.1}) is consistent with Conjecture 1 given by Zograf in \cite{Zograf2008}. This can be confirmed by comparing the Weil--Petersson volume with intersection numbers using \cite{Mirzakhani2007int}
\begin{eqnarray}
\label{volume intersection in 3.1}
V_{g,2}(b_1, b_2)= & \int_{\overline{{\cal M}}_{g,2}}{\rm exp}\big(2\pi^2\kappa_1+\frac{1}{2}(b_1^2\psi_1+b_2^2\psi_2)\big) \\ \nonumber
= & \sum_{3g-1\ge i+j\ge 0} \frac{(2\pi^2)^{3g-1-i-j}}{i!j!(3g-1-i-j)!}\, (\frac{b_1^2}{2})^i(\frac{b_2^2}{2})^j\, <\psi_1^i \, \psi_2^j \,\kappa_1^{3g-1-i-j}>,
\end{eqnarray}
\footnote{The first Miller--Morita--Mumford class $\kappa_1$ is cohomologous to $\frac{1}{2\pi^2}$ times the Weil--Petersson symplectic form $\omega$, $\kappa_1=\frac{\omega}{2\pi^2}$, owing to results in \cite{Wolpert1983, Wolpert1986}.} then setting $b_1$ and $b_2$ to zero. ($i,j$ on the right-hand side of (\ref{volume intersection in 3.1}) are non-negative integers, and their sum ranges from 0 to $3g-1$.) Zograf's conjectures in \cite{Zograf2008} for the Weil--Petersson volume when the genus $g$ is large are partially proved rigorously in \cite{MZ1999, Mirzakhani2013, MZ2011}. The fact that (\ref{Vg2 in 3.1}) is consistent with Conjecture 1 in \cite{Zograf2008} also supports our expression. 
\par We compared the precise computations of Peter Zograf of $V_{g,2}(b_1, b_2)$ up to $g=18$ \cite{Zografdata} with our expression (\ref{Vg2 in 3.1}). When $V_{g,2}(b_1, b_2)$ is expanded in $b_2^{2i}b_2^{2j}$ as 
\begin{equation}
\label{polynomial expansion of Vg2 in 3.1}
V_{g,2}(b_1,b_2)=\sum_{3g-1\ge i+j\ge 0} \, c_{gij}\, b_1^{2i}b_2^{2j},
\end{equation}
(where $c_{gij}=\frac{(2\pi^2)^{3g-1-i-j}}{i!j!(3g-1-i-j)!}\, (\frac{1}{2})^{i+j} <\psi_1^i\, \psi_2^j\, \kappa_1^{3g-1-i-j}>$ \cite{Mirzakhani2007int}) \footnote{The right-hand side of equation (\ref{polynomial expansion of Vg2 in 3.1}) is not summed over $g$. We placed $g$ in the subscript of the coefficient $c_{gij}$ to indicate that it depends on the genus $g$.}, we compared the coefficients of $b_2^{2i}b_2^{2j}$ obtained by expanding the expression (\ref{Vg2 in 3.1}) in terms of $b_2^{2i}b_2^{2j}$, which we denote by $c_{gij}^{\rm asymp.}$, with the precise coefficient computed by Zograf, which we denote by $c_{gij}^{\rm precise}$. 
\par For example, when one sets $b_1=b_2=0$, one obtains the coefficient $c_{g00}$. This corresponds to $\frac{(2\pi^2)^{3g-1}}{(3g-1)!}<\kappa_1^{3g-1}>$. We compared the constant term in (\ref{Vg2 in 3.1}) with the results by Zograf \cite{Zografdata}. The error percentages obtained from 
\begin{equation}
1-\frac{c_{g00}^{\rm precise}}{c_{g00}^{asymp.}}
\end{equation}
are within 6 \% for $2\le g\le 18$. The precision of the expression (\ref{Vg2 in 3.1}) improves as genus $g$ increases: for $6\le g\le 18$, the error percentages are less than 2\%, and for $12\le g \le 18$, the error percentages are less than 1 \%.
\par We also compared the coefficients of $b_1^{2i}b_2^{2j}$ for $0\le i,j \le 2$ (where $i$ and $j$ are not simultaneously zero) obtained from the asymptotic expression (\ref{Vg2 in 3.1}) with the results computed by Zograf \cite{Zografdata} when $g=18$. Because coefficients are symmetric under the exchange of $i$ and $j$, we only consider the case $i\ge j$ here. For all the cases $(i,j)=(1,0)$, (1,1), (2,0), (2,1), (2,2), the error percentages obtained from 
\begin{equation}
1-\frac{c_{gij}^{\rm precise}}{c_{gij}^{asymp.}}
\end{equation}
are less than 10\%. Except for the case $(i,j)=(2,2)$, the error percentages are less than 6\%. The comparison seems to suggest that genus $g$ required for a precision of agreement, say error percentage of less than 2\%, increases as $i$ and $j$ rise.

\vspace{5mm}

\par It is worth noting that the asymptotic expression (\ref{Vg2 in 3.1}) for $V_{g,2}(b_1, b_2)$ yields a leading term in $g$. There is a correction term of order $\sim 1/g$. One can see this as follows: The Weil--Petersson volume $V_{g,n}(b_1, \ldots, b_n)$ also satisfies an integral equation \cite{DoNorbury, Do2008}:
\begin{equation}
\label{integral in 3.1}
V_{g, n+1}(b_1, \ldots, b_n, 2\pi i) = \sum^n_{j=1} \int_0^{b_j} b_j V_{g,n}(b_1, \ldots, b_n) db_j.
\end{equation}
When there are two boundaries, this equation takes the following particular form:
\begin{equation}
V_{g, 2}(b_1, 2\pi i) = \int_0^{b_1} b_1 V_{g,1}(b_1) db_1.
\end{equation}
\par Integrating the result for $V_{g,1}(b_1)$ (\ref{Vg1 in 3.1}) in \cite{SSS2019}, we expect 
\begin{equation}
V_{g, 2}(b_1, 2\pi i) = \frac{4^2\, (4\pi^2)^{2g-\frac{3}{2}}}{(2\pi)^\frac{3}{2}} \Gamma(2g-\frac{3}{2})\, {\rm sinh}^2(\frac{b_1}{4}).
\end{equation}
However, expression (\ref{Vg2 in 3.1}) for $V_{g,2}(b_1, b_2)$ vanishes when $b_2=2\pi i$ \footnote{With our asymptotic expression (\ref{Vg2 in 3.1}) for $V_{g,2}(b_1, b_2)$, we have $\frac{V_{g,1}}{V_{g,2}}\sim \frac{\Gamma(2g-\frac{3}{2})}{\Gamma(2g-\frac{1}{2})}=\frac{2}{4g-3}$. Additionally, $g>>b_1$ under (\ref{condition g2 in 3.1}). Therefore, we may set $\int_0^{b_1} b_1 V_{g,1}(b_1) db_1$ to zero to the leading order in $g$. Based on this reasoning, for leading order in $g$, the vanishing of (\ref{Vg2 in 3.1}) when $b_2$ assumes the value $2\pi i$ does not suggest an inconsistency here.}. This suggests that there is a subleading correction term $\sim 1/g$ to the leading term (\ref{Vg2 in 3.1}), $\frac{4^2\, (4\pi^2)^{2g-\frac{3}{2}}}{2\pi^\frac{3}{2}} \, \Gamma(2g-\frac{3}{2})\, f(b_1, b_2)$, where $f(b_1, 2\pi i)= {\rm sinh}^2(\frac{b_1}{4})$ and $f(b_1, b_2)$ does not depend on $g$. The function $f(b_1, b_2)$ must be symmetric under exchange of $b_1$ and $b_2$. 
\par Therefore, from the above argument, we deduce the following more precise asymptotic expression for $V_{g,2}$ under (\ref{condition g2 in 3.1}):
\begin{equation}
V_{g,2}(b_1, b_2) \sim 4\sqrt{\frac{2}{\pi}}(4\pi^2)^{2g-1}\Gamma(2g-\frac{1}{2})\, \frac{{\rm sinh}(\frac{b_1}{2})}{b_1}\frac{{\rm sinh}(\frac{b_2}{2})}{b_2}+ \frac{4^2\, (4\pi^2)^{2g-\frac{3}{2}}}{(2\pi)^\frac{3}{2}} \, \Gamma(2g-\frac{3}{2})\, f(b_1, b_2).
\end{equation}
The form of the function $f(b_1, b_2)$ is left undetermined.

\subsection{Asymptotic $V_{g,n}(b_1, \ldots, b_n)$}
\label{sec3.2}
Now, we would like to estimate the asymptotic expression for general $V_{g,n}(b_1, \ldots, b_n)$ when condition (\ref{conditions in 2}) is satisfied. First, we estimate the Weil--Petersson volume with three boundaries $V_{g,3}$ for $g>>1, g>>b_1, b_2, b_3$. For this situation, $V_{g,3}$ must satisfy the following two differential equations \cite{DoNorbury, Do2008}:
\begin{eqnarray}
\label{DN equation in 3.2}
\partial_3 V_{g, 3}(b_1, b_2, 2\pi i) = & 2\pi i \cdot 2g\, V_{g,2} (b_1, b_2) \\ \nonumber
\partial^2_3 V_{g, 3}(b_1, b_2, 2\pi i) = & -(4g-2) V_{g,2}(b_1, b_2). 
\end{eqnarray}
We used the reduced equation (\ref{reduced 2nd equation in 2}) for the second equation in (\ref{DN equations in 2}) owing to the conditions $g>>1, g>> b_1, b_2, b_3$. 
\par By a similar argument to that presented in Section \ref{sec3.1}, we exploit symmetry to estimate the asymptotic expression for $V_{g,3}(b_1, b_2, b_3)$:
\begin{equation}
\label{Vg,3 in 3.2}
V_{g,3}(b_1, b_2, b_3) \sim 8\sqrt{\frac{2}{\pi}}(4\pi^2)^{2g}\Gamma(2g+\frac{1}{2}) \, \frac{{\rm sinh}(\frac{b_1}{2})}{b_1}\frac{{\rm sinh}(\frac{b_2}{2})}{b_2}\frac{{\rm sinh}(\frac{b_3}{2})}{b_3}.
\end{equation}
Assuming that $V_{g,2}(b_1, b_2)$ is given by (\ref{Vg2 in 3.1}) under (\ref{condition g2 in 3.1}), expression (\ref{Vg,3 in 3.2}) satisfies the two equations in (\ref{DN equation in 3.2}) (when $g$ tends toward infinity). Furthermore, expression (\ref{Vg,3 in 3.2}) is consistent with Conjecture 1 in \cite{Zograf2008}, as can be confirmed by setting $b_1, b_2, b_3$ in (\ref{Vg,3 in 3.2}) to zero. These results support our expression (\ref{Vg,3 in 3.2}) to some degree. 
\par Similar to the Weil--Petersson volume with two boundaries $V_{g,2}(b_1, b_2)$, $V_{g,3}(b_1, b_2, b_3)$ must satisfy the integral equation \cite{DoNorbury, Do2008}:
\begin{equation}
\label{integral Vg3 in 3.2}
V_{g, 3}(b_1, b_2, 2\pi i) = \int_0^{b_1} b_1 V_{g,2}(b_1, b_2) db_1+\int_0^{b_2} b_2 V_{g,2}(b_1, b_2) db_2.
\end{equation}
With our expression (\ref{Vg,3 in 3.2}) for $V_{g,3}(b_1, b_2, b_3)$, the left-hand side in (\ref{integral Vg3 in 3.2}) vanishes, while the right-hand side does not vanish when (\ref{Vg2 in 3.1}) is substituted into $V_{g,2}(b_1, b_2)$. This suggests that there is a subleading correction $\sim 1/g$, similar to the situation discussed in Section \ref{sec3.1}. Integrating (\ref{Vg2 in 3.1}) times $b_1$ and $b_2$, we deduce that the subleading correction is of the form
\begin{equation}
(4\pi^2)^{2g-1}\Gamma(2g-\frac{1}{2})\, f(b_1, b_2, b_3),
\end{equation}
where $f(b_1, b_2, b_3)$ is symmetric in $b_1, b_2, b_3$ and $f(b_1, b_2, b_3)$ does not depend on $g$. 

\vspace{5mm}

\par Iteration of this computation leads to the following conjecture on the asymptotic expression for $V_{g,n}(b_1, \ldots, b_n)$ under the conditions (\ref{conditions in 2}):
\begin{equation}
\label{Vgn in 3.2}
V_{g,n}(b_1, \ldots, b_n) \sim 2^n\sqrt{\frac{2}{\pi}}(4\pi^2)^{2g+n-3}\Gamma(2g+n-\frac{5}{2}) \, \prod^n_{i=1}\frac{{\rm sinh}(\frac{b_i}{2})}{b_i}.
\end{equation}
As $g$ tends toward infinity, this expression satisfies the two differential equations mentioned in Section \ref{sec2}:
\begin{eqnarray}
\label{DN for general n in 3.2}
\partial_{n+1} V_{g, n+1}(b_1, \ldots, b_n, 2\pi i) = & 2\pi i (2g-2+n)\, V_{g,n} (b_1, \ldots, b_n) \\ \nonumber
\partial^2_{n+1} V_{g, n+1}(b_1, \ldots, b_n, 2\pi i) = & -(4g-4+n) V_{g,n}(b_1, \ldots, b_n). 
\end{eqnarray}
Here, we used the reduced equation for the second equation in (\ref{DN for general n in 3.2}) owing to the conditions (\ref{conditions in 2}), as previously explained in Section \ref{sec2}. Furthermore, setting $b_1, b_2, \ldots, b_n$ to zero, the consistency of expression (\ref{Vgn in 3.2}) with Conjecture 1 in \cite{Zograf2008} is confirmed. 
\par $V_{g,n}(b_1, \ldots, b_n)$ must also satisfy the integral equation (\ref{integral in 3.1}) \cite{DoNorbury, Do2008}. This suggests that there is a subleading correction $\sim 1/g$ to expression (\ref{Vgn in 3.2}), similar to $V_{g,2}(b_1, b_2)$ and $V_{g,3}(b_1, b_2, b_3)$, which we have already discussed.

\vspace{5mm}

\par We make a remark here that the asymptotic Weil--Petersson volumes can also be evaluated from the matrix integral perspective. 
\par Correlation function of resolvents $<R(E_1)R(E_2)\ldots R(E_n)>_c$ is expanded as follows \cite{SSS2019}:
\begin{equation}
<R(E_1)R(E_2)\ldots R(E_n)>_c \simeq \sum_{g=0}^{\infty} e^{(-2g+2-n)S_0}R_{g,n}(E_1, \ldots, E_n).
\end{equation}
Functions $W_{g,n}$ \footnote{The functions $W_{g,n}$ satisfy a recursion relation \cite{Eynard2004} that corresponds to Mirzakhani's recursion relation via Laplace transform \cite{Eynard2007}.} are defined as \cite{Eynard2014, SSS2019}:
\begin{equation}
\label{definition W in 3.2}
W_{g,n}(z_1, \ldots, z_n) = 2^n (-1)^n \cdot z_1\ldots z_n \cdot R_{g,n}(-z_1^2, -z_2^2, \ldots, -z_n^2).
\end{equation}
Then, the following equality holds \cite{Eynard2007}:
\begin{equation}
\label{equality W and volume in 3.2}
W_{g,n}(z_1, \ldots, z_n) = \prod_{i=1}^n \int_0^{\infty} b_i db_i e^{-b_i z_i} V_{g,n}(b_1, \ldots, b_n)
\end{equation}
Using the definition of $W_{g,n}$ (\ref{definition W in 3.2}), the equality (\ref{equality W and volume in 3.2}) can be rewritten as follows:
\begin{equation}
\label{equality R and volume in 3.2}
R_{g,n}(-z_1^2, -z_2^2, \ldots, -z_n^2)= (-1)^n \frac{1}{2^n\, z_1\ldots z_n }\prod_{i=1}^n \int_0^{\infty} b_i db_i e^{-b_i z_i} V_{g,n}(b_1, \ldots, b_n).
\end{equation}
Multiplying the two sides in the equation (\ref{equality R and volume in 3.2}) by $ e^{(-2g+2-n)S_0}$ and summing up over genus $g$, the following equality is obtained:
\begin{equation}
\label{correlator and volume in 3.2}
<R(-z_1^2)R(-z_2^2)\ldots R(-z_n^2)>_c = (-1)^n \frac{1}{2^n\, z_1\ldots z_n }\prod_{i=1}^n \int_0^{\infty} b_i db_i e^{-b_i z_i} \sum_{g} e^{(-2g+2-n)S_0}\, V_{g,n}(b_1, \ldots, b_n).
\end{equation}
\par The case with one boundary, $n=1$, was discussed, and the volume $V_{g,1}(b)$ was evaluated from the expectation value of the resolvent $<R(-z^2)>$ in \cite{SSS2019}. The equation (\ref{correlator and volume in 3.2}) yields straightforward generalization to an arbitrary number of boundaries. We expect that by using an argument similar to that given in \cite{SSS2019}, computing correlation function of the resolvents and applying inverse Laplace transform, the Weil--Petersson volume $V_{g,n}(b_1, \ldots, b_n)$ can also be estimated from the matrix integral perspective.

\section{Asymptotic behavior of the Weil--Petersson volume \\ $V_{g,n}(b_1, \ldots, b_n)$ with large $b$}
\label{sec4}
\par Mirzakhani proved \cite{Mirzakhani2007} that the Weil--Petersson volume $V_{g,n}(b_1, \ldots, b_n)$ is a polynomial in $b$ given by
\begin{equation}
\label{volume polynomial in 4}
V_{g,n}(b_1, \ldots, b_n)= \sum_{|\alpha|\le 3g-3+n} c_g(\alpha)\, b^{2\alpha},
\end{equation}
where $b$ on the right-hand side represents $b=(b_1, b_2, \ldots, b_n)$, $\alpha=(\alpha_1, \alpha_2, \ldots, \alpha_n)$, and $\alpha_i$, $i=1, \ldots, n$ are non-negative integers. $b^{2\alpha}$ is defined as $b^{2\alpha}=b_1^{2\alpha_1}\, b_2^{2\alpha_2}\, \ldots \, b_n^{2\alpha_n}$. The coefficient $c_g(\alpha)$ is positive: $c_g(\alpha)>0$, and $c_g(\alpha)$ is in $\pi^{6g-6+2n-2|\alpha|}\cdot \Q$. $|\alpha|$ is defined as $|\alpha|=\sum^n_{i=1}\alpha_i$ in (\ref{volume polynomial in 4}). \footnote{The coefficients $c_g(\alpha)$ in (\ref{volume polynomial in 4}) are expressed as intersection numbers of cohomology classes on the moduli of Riemann surfaces in \cite{Mirzakhani2007int}. When $n=2$, the equation (\ref{volume polynomial in 4}) corresponds to the equations (\ref{volume intersection in 3.1}) and (\ref{polynomial expansion of Vg2 in 3.1}).} 
\par With regard to the theorem of Mirzakhani, it is worth making a remark about the Weil--Petersson volume in the limit where the boundary lengths $b$ become large, i.e., $b>>g>>1$. It follows from Mirzakhani's theorem that in the large $b$ regime ($b_1, \ldots, b_n >>g>>1$), the highest-order terms in the polynomial (\ref{volume polynomial in 4}) are dominant. Therefore, in the regime $b_1, \ldots, b_n >>g>>1$, the Weil--Petersson volume should be approximated by the sum of these highest-order terms in $b$:
\begin{equation}
\label{asymptotic volume in 4}
V_{g,n}(b_1, \ldots, b_n) \simeq \sum_{|\alpha|=3g-3+n} c_g(\alpha)\, b^{2\alpha},
\end{equation}
where the coefficients $c_g(\alpha)$ with $|\alpha|=3g-3+n$ are positive and are in $\Q$. 
\par However, the method discussed in this report, which employs partial differential equations (\ref{DN equations in 2}) to estimate the asymptotic Weil--Petersson volume, does not apply in the large $b$ regime to determine the coefficients in (\ref{asymptotic volume in 4}). This is owing to the fact that the conditions $b_1, \ldots, b_n >> g >>1$ are imposed on $b$. 
\par It might be interesting to consider if there is a method for estimating the asymptotic expression for the Weil--Petersson volume $V_{g,n}(b_1, \ldots, b_n)$ in the regime $b_1, \ldots, b_n >> g >>1$, from the volume $V_{g,n-1}(b_1, \ldots, b_{n-1})$. This is equivalent to estimating the coefficients $c_g(\alpha)$ in (\ref{asymptotic volume in 4}) for asymptotic $V_{g,n}$, when the coefficients $c_g(\alpha)$ are known for asymptotic $V_{g,m}$, $m<n$, in this regime. 

\vspace{5mm}

\par The asymptotic Weil--Petersson volume for any genus for the case of one boundary length becoming large is computed in Appendix A of \cite{Maxfield202006}. This corresponds to the situation $b_1 >>g$, $b_1>>1$. For this situation, the term with $\alpha=(3g-3+n, 0, \ldots, 0)$ yields a unique leading term, and the coefficient $c_g(3g-3+n, 0, \ldots, 0)$ of the leading term is given by $\frac{1}{(24)^g\cdot 2^{3g-3+n}\cdot g!(3g-3+n)!}$ \cite{Maxfield202006}.

\section{Concluding remarks and open problems}
\label{sec5}
Herein, we utilized partial differential equations satisfied by the Weil--Petersson volume \cite{DoNorbury, Do2008} to estimate the asymptotic Weil--Petersson volumes $V_{g,2}(b_1, b_2)$ and $V_{g,3}(b_1, b_2, b_3)$ for large genus $g$ ($g >>1$ and $g >> b_1, b_2, b_3$). We also conjectured the asymptotic expression for the volume $V_{g,n}(b_1, \ldots, b_n)$ for general $n$, when the genus $g$ is large ($g >>1$ and $g>> b_1, \ldots, b_n$). The obtained asymptotic expressions satisfy the partial differential equations deduced in \cite{DoNorbury, Do2008} to leading order in $g$. We also confirmed that, when the $b_i$ all vanish, our asymptotic expressions for the volumes ($V_{g,2}(b_1, b_2)$ with two boundaries, $V_{g,3}(b_1, b_2, b_3)$ with three boundaries, and $V_{g,n}(b_1, \ldots, b_n)$ in general) are all consistent with Conjecture 1 for the asymptotic Weil--Petersson volumes specified by Zograf in \cite{Zograf2008}. We also compared our asymptotic expression (\ref{Vg2 in 3.1}) for $V_{g,2}(b_1, b_2)$ with the precise computational results by Zograf \cite{Zografdata}. The comparison showed good agreement.  

\par The Weil--Petersson volumes yield the intersection numbers of cohomology classes on the moduli of Riemann surfaces \cite{Mirzakhani2007int}. For example, Mirzakhani proved \cite{Mirzakhani2007int} Witten's conjecture \cite{Witten1990} \footnote{Witten's conjecture was proved in \cite{Kontsevich1992}. Proofs of Witten's conjecture can also be found in \cite{Okounkov2000, Kazarian2007}.} by relating the intersection numbers of cohomology classes on the moduli of Riemann surfaces to the coefficients of the Weil--Petersson volume in (\ref{volume polynomial in 4}). The asymptotic expressions for the Weil--Petersson volumes obtained in this note provide information on the intersection numbers of certain cohomology classes on the moduli of Riemann surfaces. It might be interesting to compare the known intersection numbers of cohomology classes with our results. Physically, they include the correlation functions of 2d topological gravity \cite{Dijkgraaf2018}.

\par The asymptotic Weil--Petersson volumes $V_{g,0}$ and $V_{g,1}(b)$ were predicted in \cite{SSS2019} in the context of bosonic JT gravity, by using the density of eigenvalues via topological recursion. As discussed in \cite{SSS2019}, their method of using topological recursion \footnote{Discussions of a related approach can be found, e.g, in \cite{Marino2006, Marino2007, Marino2012}.} to predict asymptotic Weil--Petersson volumes may extend to general $V_{g,n}(b_1, \ldots, b_n)$. If so, it might also be interesting to study whether our expressions agree with the results obtained from the approach of topological recursion.
\par If the method of topological recursion succeeds in evaluating the asymptotic Weil--Petersson volume $V_{g,n}(b_1, \ldots, b_n)$ in the regime $b_1, \ldots, b_n >> g >> 1$, the saddle-point approximation should yield the approximation of the Weil--Petersson volume $V_{g,n}(b_1, \ldots, b_n)$ in the form (\ref{asymptotic volume in 4}) discussed in Section \ref{sec4}. This provides a non-trivial check of the result given by the topological recursion, if it is applied to evaluate the Weil--Petersson volume $V_{g,n}(b_1, \ldots, b_n)$ for general $n$ in the regime $b_1, \ldots, b_n >> g >> 1$.

\section*{Acknowledgments}

We are grateful to Peter Zograf for providing us the computational results of the volume $V_{g,2}(b_1, b_2)$. We would like to thank Kazuhiro Sakai for discussions.

\end{document}